\documentclass[twocolumn,tighten,times,astrosymb,floatfix]{aastex631}

\usepackage{
	amsmath,
	amssymb,
	amsthm,
	booktabs,
	epsfig,
	graphicx,
	grffile,
	import,
	morefloats,
	microtype,
	natbib,
	subfigure,
	ulem,
	url,
	verbatim,
	wasysym,
	xcolor,
	xspace,
}
\usepackage{savesym}
\savesymbol{tablenum}
\usepackage{siunitx}
\restoresymbol{SIX}{tablenum}
\usepackage[version=4]{mhchem}

\DeclareSIUnit{\jy}{Jy}
\DeclareSIUnit{\beam}{beam}
\DeclareSIUnit{\kms}{\kilo\meter\per\second}

\shorttitle{Beta Pic Time Variability}
\shortauthors{Chen et al.}

\begin{document}

\title{\Large MIRI MRS Observations of Beta Pictoris II. The Spectroscopic Case for a Recent Giant Collision
}

\correspondingauthor{Christine H. Chen}
\email{cchen@stsci.edu}

\author[0000-0002-8382-0447]{Christine H. Chen}
\affiliation{Space Telescope Science Institute, 3700 San Martin Drive, Baltimore, MD 21218, USA }
\affiliation{William H. Miller III Department of Physics and Astronomy, John's Hopkins University, 3400 N. Charles Street, Baltimore, MD 21218, USA}

\author[0000-0001-9352-0248]{Cicero X. Lu}
\affiliation{Gemini Observatory/NSF’s NOIRLab, 670N. A’ohokuPlace, Hilo, HI 96720, USA}
\affiliation{Science Fellow, Gemini North}

\author[0000-0002-5885-5779]{Kadin Worthen}
\affiliation{William H. Miller III Department of Physics and Astronomy, John's Hopkins University, 3400 N. Charles Street, Baltimore, MD 21218, USA}

\author[0000-0002-9402-186X]{David R. Law}
\affiliation{Space Telescope Science Institute, 3700 San Martin Drive, Baltimore, MD 21218, USA }

\author[0000-0001-9855-8261]{B. A. Sargent}
\affiliation{Space Telescope Science Institute, 3700 San Martin Drive, Baltimore, MD 21218, USA }
\affiliation{William H. Miller III Department of Physics and Astronomy, John's Hopkins University, 3400 N. Charles Street, Baltimore, MD 21218, USA}

\author[0000-0001-9504-8426]{Amaya Moro-Martin}
\affiliation{Space Telescope Science Institute, 3700 San Martin Drive, Baltimore, MD 21218, USA }

\author[0000-0003-4520-1044]{G.~C.\ Sloan}
\affiliation{Space Telescope Science Institute, 3700 San Martin Drive, Baltimore, MD 21218, USA }
\affiliation{Department of Physics and Astronomy, University of North Carolina, Chapel Hill, NC 27599-3255, USA}

\author[0000-0002-9548-1526]{Carey M. Lisse}
\affiliation{Johns Hopkins University Applied Physics Laboratory, 11100 Johns Hopkins Rd, Laurel, MD 20723, USA}

\author[0000-0001-8302-0530]{Dan M. Watson}
\affiliation{Department of Physics and Astronomy, University of Rochester, 500 Wilson Blvd, Rochester, NY 14627, USA}

\author[0000-0001-8627-0404]{Julien H. Girard}
\affiliation{Space Telescope Science Institute, 3700 San Martin Drive, Baltimore, MD 21218, USA }

\author[0009-0008-5865-5831]{Yiwei Chai}
\affiliation{William H. Miller III Department of Physics and Astronomy, John's Hopkins University, 3400 N. Charles Street, Baltimore, MD 21218, USA}

\author[0000-0003-4653-6161]{Dean C. Hines}
\affiliation{Space Telescope Science Institute, 3700 San Martin Drive, Baltimore, MD 21218, USA }

\author[0000-0003-2769-0438]{Jens Kammerer}
\affiliation{European Southern Observatory, Karl-Schwarzschild-Straße 2, 85748 Garching, Germany}
\affiliation{Space Telescope Science Institute, 3700 San Martin Drive, Baltimore, MD 21218, USA }

\author{Alexis Li}
\affiliation{William H. Miller III Department of Physics and Astronomy, John's Hopkins University, 3400 N. Charles Street, Baltimore, MD 21218, USA}

\author[0000-0002-3191-8151]{Marshall Perrin}
\affiliation{Space Telescope Science Institute, 3700 San Martin Drive, Baltimore, MD 21218, USA }

\author[0000-0003-3818-408X]{Laurent Pueyo}
\affiliation{Space Telescope Science Institute, 3700 San Martin Drive, Baltimore, MD 21218, USA }

\author[0000-0002-4388-6417]{Isabel Rebollido}
\affiliation{European Space Agency (ESA), European Space Astronomy Centre (ESAC), Camino Bajo del Castillo s/n, 28692 Villanueva de la Ca\~nada, Madrid, Spain}

\author[0000-0002-2805-7338]{Karl R. Stapelfeldt}
\affiliation{NASA Jet Propulsion Laboratory, 4800 Oak Grove Drive, La Ca\~nada, CA 91011, USA}

\author{Christopher Stark}
\affiliation{NASA Goddard Space Flight Center, Exoplanets $\&$ Stellar Astrophysics Laboratory, Code 667, Greenbelt, MD 20771, USA}

\author[0000-0003-4990-189X]{Michael W. Werner}
\affiliation{NASA Jet Propulsion Laboratory, 4800 Oak Grove Drive, La Ca\~nada, CA 91011, USA}

\begin{abstract}
Modeling observations of the archetypal debris disk around $\beta$ Pic, obtained in 2023 January with the MIRI MRS on board \emph{JWST}, reveals significant differences compared with that obtained with the IRS on board \emph{Spitzer}. The bright 5 - 15 $\mu$m continuum excess modeled using a $\sim$600 K black body has disappeared. The previously prominent 18 and 23 $\mu$m crystalline forsterite emission features, arising from cold dust ($\sim$100 K) in the Rayleigh limit, have disappeared and been replaced by very weak features arising from the hotter 500 K dust population. Finally, the shape of the 10 $\mu$m silicate feature has changed, consistent with a shift in the temperature of the warm dust population from $\sim$300 K to $\sim$500 K and an increase in the crystalline fraction of the warm, silicate dust. Stellar radiation pressure may have blown both the hot and the cold crystalline dust particles observed in the \emph{Spitzer} spectra out of the planetary system during the intervening 20 years between the \emph{Spitzer} and \emph{JWST} observations. These results indicate that the $\beta$ Pic system has a dynamic circumstellar environment, and that periods of enhanced collisions can create large clouds of dust that sweep through the planetary system.
\end{abstract}

\keywords{
    planet formation ---
    debris disks ---
    circumstellar matter
}


\section{Introduction}

The planetary system around the nearby (19.4 pc), $\sim$20 Myr old \citep{miret20}, A6V star $\beta$ Pictoris is one of the most well studied young, planetary systems. The star hosts a bright, extended (up to 400 au) debris disk that has been imaged in detail both in scattered light \citep{golimowski06,heap00,smith84} and thermal emission \citep{dent14,telesco05, rebollido24} and two $\sim$10-12 M$_{Jup}$ companions, one at 10.0 au and another at 2.7 au \citep{lagrange19,lagrange10}. $\beta$ Pic's debris disk is noteable because it was one of the first to be discovered. Routine calibration observations using the \emph{Infrared Astronomical Satellite (IRAS)} revealed strong, unresolved mid- to far-infrared excess at 12, 25, 60 and 100 $\mu$m, consistent with the presence of circumstellar dust \citep{aumann84}. Among debris disks, $\beta$ Pic's debris disk has a relatively high fractional infrared luminosity ($L_{IR}/L_{*}$ $\sim$ 10$^{-3}$), consistent with its young age \citep{lagrange00}. Estimates of the Poynting-Robertson Drag and collisional lifetimes of the circumstellar dust grains indicate that the dust particles in the inner disk ($\sim$40 au) are destroyed in collisions with other dust particles before they can spiral into the orbit center \citep{backman93}. High resolution visual spectroscopy has revealed the presence of red-shifted, absorption features that are variable on timescales as short as hours, and have been attributed to infalling exocomets \citep{beust98,ferlet87}. In contrast, little is known about the time variability of the circumstellar dust.

During the past decade, \emph{Spitzer Space Telescope}, \emph{Stratospheric Observatory for Infrared Astronomy (SOFIA)}, and \emph{Wide-field Infrared Survey Explorer (WISE)} monitoring of "extreme" debris disks with fractional infrared luminosities ($L_{IR}/L_{*}$ $>$ 0.01) has revealed time variable behavior over periods as short as months \citep{moor21,meng15}. For example, \emph{Spitzer} IRAC 3.6 and 4.5 $\mu$m monitoring has revealed both short-term (weekly to monthly) and long-term (yearly) variability on the order of 10-80\% in the infrared excess for ID 8 in NGC 2547 and P1121 in M47 \citep{su19}. Since these objects have estimated ages 35 Myr and 80 Myr, respectively, they are expected to be forming terrestrial planets. The observations can be explained if a giant collision created an initially optically thick dust cloud containing vapor condensates that continued to collide on orbital timescales. However, not all extreme debris disks exhibit periodic behavior. In contrast, the 4-5 Myr old star HD 166191 and the 80 Myr old v488 Per (in $\alpha$ Per) were relatively quiescent for $\sim$3 years before showing large changes in their excess fluxes \citep{su22, rieke21}. The timescales and changes in the HD 166191 and v488 Per infrared excesses are consistent with the collisional destruction of a single Vesta-sized object within the terrestrial planet zone. 

High angular resolution imaging of the dust and gas in the $\beta$ Pic debris disk has revealed fine structure that may have been generated by recent collisions. High angular resolution, ground-based, mid-infrared imaging using Gemini South T-ReCS discovered a brightness peak on the southwest side of the disk at 52 au that was attributed to a dust clump created in a recent collision \citep{telesco05}. Long term monitoring of this structure from similar facilities has searched for orbital motion but yielded discrepant hence inconclusive results \citep{skaf23,han23,li12}. In tandem, high angular resolution millimeter interferometry has revealed an enhancement in \ion{C}{1} and CO also on the southwest side of the disk at slight larger distances from the star, 84 au \citep{Cataldi+18,dent14}. The origin of the CO clump has not yet been firmly established. It may be generated by steady-state collisions among planetesimals trapped in a mean motion resonance with an undetected planet or the by product of a recent collision. Within the past year, \emph{JWST} MIRI Coronagraphic imaging at 15.5 $\mu$m has discovered a new structure on the southwest side of the disk that has been dubbed the "cat's tail" in addition to the fork on the northeast side of the disk that is related to the inclined inner disk \citep{rebollido24}. The color of the cat's tail is significantly bluer than the main disk, consistent with near blow out, fluffy grains composed of organic material. \cite{rebollido24} hypothesized that the cat's tail, along with the fork, may be explained by a single family of collisions at 85 au on the southwest side of the disk, the most recent of which occurred $\sim$150-200 years ago.

Solid-state spectroscopy has been used to characterize the $\beta$ Pic dust properties (including grain composition, size, and crystalline fraction), initially beginning with ground based observations of the 10 $\mu$m silicate emission feature \citep{li12,skinner92} and expanding to solid-state features at longer wavelengths. High angular resolution, spatially resolved measurements of the 10 $\mu$m silicate emission feature obtained using Subaru COMICS in December 2003 indicated that the small, amorphous grains were not cospatial with the crystalline forsterite, concentrated at the location of the star and consistent with thermal annealing. Instead, the small grains appeared to be located in discrete rings at 6, 16, and 30 au \citep{okamoto04}. In 2004 and 2005, lower angular resolution, higher SNR \emph{Spitzer} IRS observations measured silicate emission features at 18, 23, 28, and 33 $\mu$m for the first time \citep{lu22,chen07}. In June 2010, \emph{Herschel} PACS observations measured the 69 $\mu$m crystalline forsterite emission feature; detailed fitting indicated that the cold dust was extremely Magnesium-rich, Fe/(Fe+Mg)=0.01 \citep{devries12}. A more recent analysis of \emph{Spitzer} IRS observations indicated that the 20 - 30 $\mu$m features were also generated by cold ($\sim$100 K), Magnesium-rich Forsterite (Fe/(Fe+Mg)=0.01). Comparison of the 10, 18, 23, and 33 $\mu$m silicate emission features indicated gradients in the dust grain composition with $\sim$300 K dust in the terrestrial planet zone being more amorphous, Iron-rich, and regular than $\sim$100 K dust in the outer planetary system \citep{lu22}. 

In January 2023, \emph{JWST} observed $\beta$ Pic using the Mid-infrared Instrument's (MIRI's) Medium Resolution Spectrograph (MRS), almost eighteen years after the last Spitzer IRS spectrum was taken. As expected, the \emph{JWST} MIRI observations have provided spatially resolved spectroscopic measurements of the planetary system, including a mid-infrared spectrum of planetary mass "b" companion and a detection of water in its atmosphere for the first time. However, these observations have also revealed surprises about the well studied archetypal disk. For example, the 18 and 23 $\mu$m crystalline forsterite emission features so prominently detected in 2004-5 now appear to be missing and a new gas species, \ion{Ar}{2}, has been detected and spatially resolved (\cite{worthen24}, hereafter Paper I). In this paper, we present a new extraction of the 2023 $\beta$ Pic \emph{JWST} MIRI spectrum, using an aperture commensurate with that used to extract the 2004-5 \emph{Spitzer} IRS spectrum, to facilitate comparison of the underlying dust properties. In Section 2, we describe the \emph{Spitzer} and \emph{JWST} observations and data reduction. In Section 3, we present silicate modeling of the solid-state features detected in the \emph{JWST} spectrum, indicating that a large mass of dust has disappeared between 2005 and 2023. In Section 4, we discuss the origin of the mass loss and the its possible implications for the planetary mass companions in the system. In Section 5, we present our conclusions.

\section{Observations}

Several mid-infrared observations of the $\beta$ Pic planetary system have been obtained in the decades since the first detection of its infrared excess using the \emph{IRAS}. In general, absolute photometric calibration and the calibration of the shape of mid-infrared spectra are challenging from the ground because the night sky is bright at 10-20 $\mu$m and observing conditions change rapidly. For example, the absolute photometric calibration of ground-based, mid-infrared photometry is typically $\sim$5-10\% \citep{cohen99} while that of \emph{Spitzer} IRAC and MIPS photometry were typically $\sim$2-3\% \citep{engelbracht07,reach05}. Unfortunately, $\beta$ Pic is substantially brighter than the \emph{WISE} and \emph{NEOWISE} saturation limits: 8.1, 6.7, 3.8, and -0.4 mag in Bands 1, 2, 3, and 4, respectively \citep{cutri13}. Thus, the photometry extracted from individual semi-annual  epochs has typical calibration uncertainties, $\sim$1.8\% and $\sim$2.5\% in Bands 1 and 2, respectively \citep{meisner23}, too large to detect and characterize time-variable behavior. Recent \emph{JWST} MIRI observations of TW Hya indicate that the \emph{Spitzer} IRS and \emph{JWST} MIRI spectra of this source are consistent \citep{Henning+24}. As a result, we focused our analysis on the comparison of well calibrated, high signal:noise \emph{Spitzer} IRS and \emph{JWST} MIRI spectra.

\begin{deluxetable*}{lccccccccc}
\tablecaption{Published \emph{Spitzer} IRS Observations of $\beta$ Pic\label{tab:irs}}
\tablehead{
\colhead{Date} & \colhead{Order} & \colhead{Wavelength} & \colhead{Mode} & \colhead{AOR Key} & \colhead{$\#$ Pointings} & Pointing Extracted & \colhead{Slit Size} & \colhead{Plate Scale} & \colhead{References} \\
\omit & \omit & \colhead{($\mu$m)} & \omit & \omit & \omit & \omit & \omit & \colhead{(arcsec pix$^{-1}$)} & \omit }
\startdata
2003 Dec 15 & SH & 9.9 - 19.6 & Mapping & 4879616 & 3 & 2 & 11.3$\arcsec$$\times$4.7$\arcsec$ & 2.3 & 1 \\
2004 Feb 04 & LL1 & 19.9 - 38.0 & Staring & 9016064 & 2 & 1, 2 & 168$\arcsec$$\times$10.5$\arcsec$ & 5.1 & 2, 3 \\
2004 Mar 04 & LH & 18.7 - 37.2 & Mapping & 4876800 & 3 & 2 & 22.3$\arcsec$$\times$11.1$\arcsec$ & 4.5 & 1 \\
2004 Nov 16 & SL1 & 7.4 - 14.5 & Mapping & 8972544 & 11 & 6 & 57$\arcsec$$\times$3.6$\arcsec$ & 1.8 & 1, 2, 3 \\
2004 Nov 16 & SL2 & 5.2 - 7.7 & Mapping & 9872288 & 7 & 4 & 57$\arcsec$$\times$3.6$\arcsec$ & 1.8 & 1, 2, 3 \\
2005 Feb 09 & LL2 & 13.9 - 21.3 & Mapping & 9016832 & 5 & 3 & 168$\arcsec$$\times$10.5$\arcsec$ & 5.1 & 2, 3 \\
\enddata
\tablerefs{(1) \cite{chen07}, (2) \cite{lu22}, (3) This work}
\end{deluxetable*}

\subsection{Spitzer IRS}
The $\beta$ Pic debris disk was observed extensively using the IRS \citep{houck04} during the \emph{Spitzer} cryogenic mission \citep{werner04}. The $\beta$ Pic observations were part of the Fabulous 4 Debris Disks program (PI Werner, Program ID 90), a collaboration amongst the \emph{Spitzer} GTOs to study the four brightest debris disks discovered using \emph{IRAS} ($\beta$ Pic, $\epsilon$ Eri Vega, Fomalhaut). As a result, the target and the observations were reserved early in the mission. \emph{Spitzer} launched on August 25, 2003 and began regular operations a little over three months later on December 1, 2003. All of the $\beta$ Pic data were obtained within the first 15 months of regular operations. 

A subset of the \emph{Spitzer} IRS observations have already appeared in the published literature (see Table~\ref{tab:irs}). Spectral mapping mode observations using the Short-Low (SL; 5.2 - 14 $\mu$m; $\lambda/\Delta \lambda \sim$ 90), Short-High (SH; 9.9 - 19.6 $\mu$m; $\lambda/\Delta \lambda \sim$ 600), and Long-High (SH; 18.7 - 37.2 $\mu$m; $\lambda/\Delta \lambda \sim$ 600) modules were published using a full slit extraction \citep{chen07}. These observations revealed the 23, 28, and 33 $\mu$m silicate emission features for the first time, showing that the infrared continuum emission was approximately consistent with scattered light disk models. They were later folded into a joint analysis with a \emph{Herschel} PACS measurement of the 69 $\mu$m forsterite feature \citep{devries12}. More recently, staring mode observations using the Long-Low (LL; 13.9 - 38 $\mu$m; $\lambda/\Delta \lambda \sim$ 90) module were published along with a recalibrated version of the SL spectra \citep{lu22}. The updated spectra were extracted using Advanced Optimal (AdOpt) Extraction and the full \emph{Spitzer} archive to calibrate the data. In their analysis, \cite{lu22} discovered an unresolved 5 $\mu$m excess and a new 18 $\mu$m silicate emission feature that required the presence of small, cold, magesium-rich forsterite grains. They showed that the SL-LL spectrum appeared significantly different from the previously extracted spectrum and attributed the differences to improved calibration and spectral extraction techniques.

In our current analysis, we focus on the SL and LL observations and reduction presented in \cite{lu22}. The SH (11.3$\arcsec$x4.7$\arcsec$) and LH (22.3$\arcsec$x11.1$\arcsec$) slits are extremely small, compared to the size of the diffraction limited PSFs, with little additional area in which to measure the background sky. Beginning in Cycle 2, the \emph{Spitzer} Science Center recommended obtaining dedicated background observations for all SH and LH observations to enable empirical background subtraction. Since the $\beta$ Pic observations were planned before the mission launched, they did not include background observations. As a result, the full slit extraction published in \cite{chen07} includes not only the emission from the disk but also the background. Since $\beta$ Pic was expected to be bright compared to the mid-infrared astrophysical background, the background contribution to the $\beta$ Pic spectra was expected to be small.

In contrast, the SL and LL observations provided an opportunity for empirical background subtraction because the observations were made in both the 1st and 2nd orders. The IRS had two separate SL slits (each 57$\arcsec$x3.6$\arcsec$) to observe SL1 and SL2. As a result, the SL1 slit observed the sky when then SL2 slit observed an unresolved point source and vice versa. Similarly, the IRS had two separate LL slits (each 168$\arcsec$x10.5$\arcsec$) to observe LL1 and LL2. Similarly, the LL1 slit observed the background sky when then LL2 slit observed an unresolved point source and vice versa. Indeed, \cite{lu22} used this background subtraction technique to background subtract the data at the detector level before extracting the spectra. Finally, they flux calibrated the combined SL and LL spectrum using a measurement of the MIPS 24 $\mu$m flux of the unresolved central point source.

In our analysis, we used one-dimensional SL and LL spectra extracted by \cite{lu22}. \cite{lu22} extracted their spectra using Advanced Optimal Extraction (AdOpt) that weights pixel values by their signal-to-noise ratios (SNRs) to maximize the SNR of the extracted spectra from the unresolved central point source. They found that $\beta$ Pic was well centered in the slit for the SL2, SL1, and LL2 mapping mode observations. However, they also found that the star was not well centered in the slit in the LL1 staring mode observations, resulting in mild fringing in the one-dimensional extracted spectrum. To correct the fringing, they applied a Relative Spectral Response Function (RSRF) to the LL1 observations, derived from observations of $\xi$ Dra with a similar mispointing. For additional details on the reduction of the \emph{Spitzer} IRS observations, please see \cite{lu22}. The calibrated \emph{Spitzer} IRS spectra contain flux from both the stellar host star and the circumstellar dust. We used the photosphere subtracted spectrum from \cite{lu22} to quantify the thermal infrared excess from the circumstellar dust. \cite{lu22} modeled the stellar photosphere using Kurucz stellar atmosphere model with $T_{*}$ = 8000 K, $\log g$ = 4.0, and stellar metallicity, normalized to fit the stellar ultra violet through near-infrared photometry and spectra. 

\begin{deluxetable*}{ccccccc}
\tablecaption{\emph{JWST} MIRI MRS $\beta$ Pic Observing Sequence\label{tab:mrs}}
\tablehead{
\colhead{Date} & \colhead{Gratings} & \colhead{Wavelength} & \colhead{Exposure} & \colhead{Object} & \colhead{Dither Pattern} & \colhead{References} \\
\omit & \omit & \colhead{($\mu$m)} & \omit & \omit & \omit & \omit
}
\startdata
2023 Jan 11 & A, B, C & 5 - 28 & 1 & Background & 2-pt Extended Source & 1, 2\\
2023 Jan 11 & A, B, C & 5 - 28 & 2 & Flanking Field & 4-pt Point Source & N/A \\
2023 Jan 11 & A, B, C & 5 - 28 & 3 & Host Star & 4-pt Point Source & 1, 2\\
2023 Jan 11 & A, B, C & 5 - 28 & 4 & N Car & 4-pt Point Source & 1, 2\\
\enddata
\tablerefs{(1) \cite{worthen24}, (2) This work}
\end{deluxetable*}

\begin{figure*}
    \centering
    \plottwo{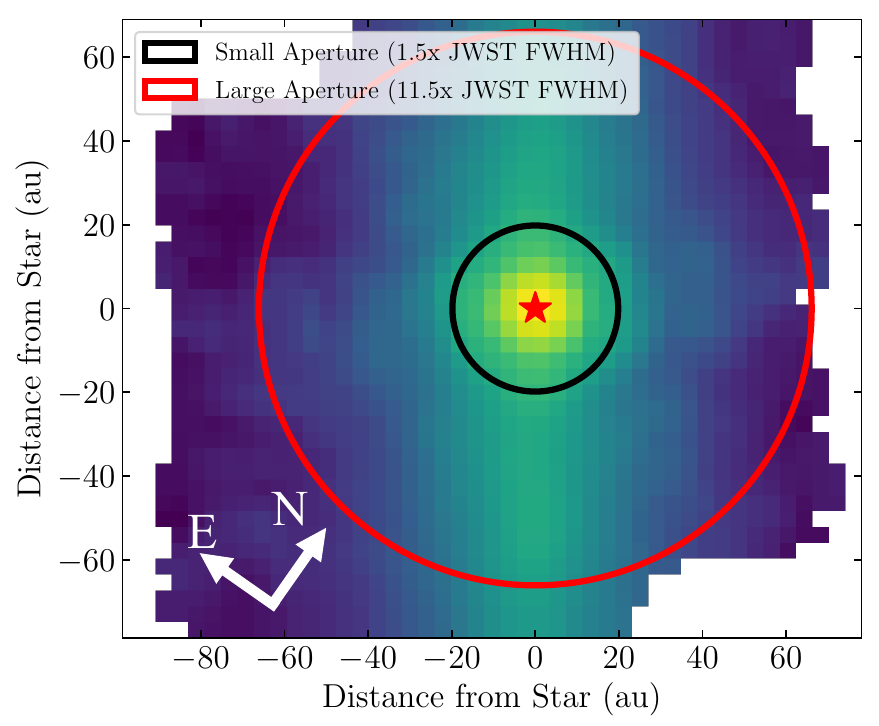}{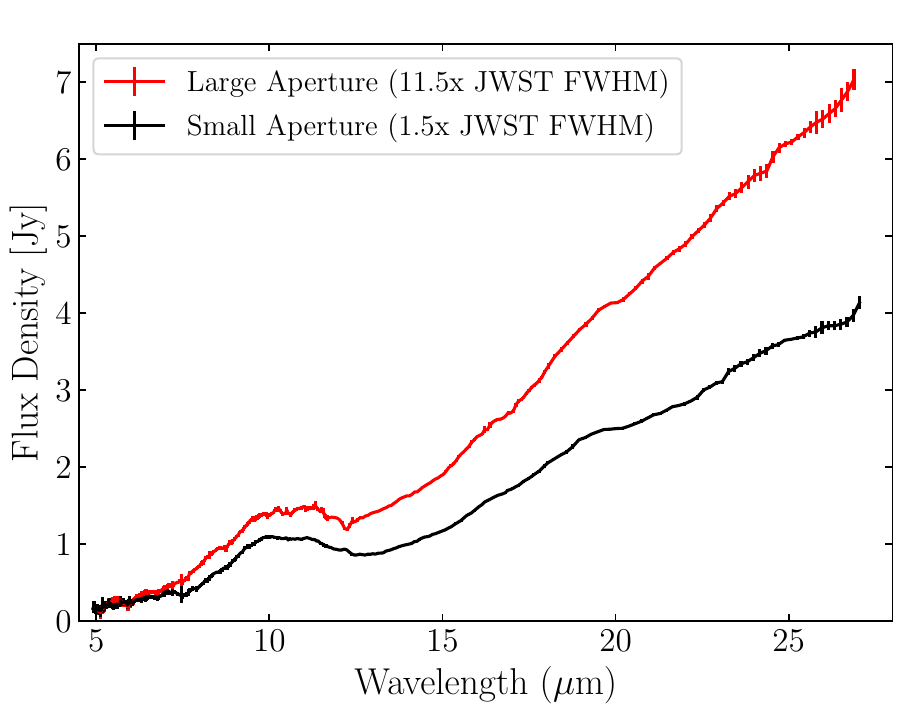}
    \caption{(Left) Slice of the Channel 3C data cube at 15.4 $\text{\textmu}$m displayed using a log stretch. The along-slice orientation is horizontal in this image. The location of host star is marked with a red star. In the instrument frame, the thermal emission from the debris disk is viewed as a vertical band of emission. Overlaid are a black circle, showing the extraction aperture with radius 1.5$\times$ the JWST FWHM, and a red circle, showing the extraction aperture equal to the angular resolution of Spitzer at this wavelength. (Right) The photosphere subtracted, MRS spectrum, extracted from an aperture with a radius equal to 1.5$\times$ the JWST PSF FWHM (black) and an aperture with a radius equal to the angular resolution of Spitzer (red).}
    \label{fig:Aperture_compare}
\end{figure*}

\subsection{JWST MIRI MRS}
The $\beta$ Pic debris disk has been observed once with \emph{JWST} using the MIRI MRS thus far. Like the \emph{Spitzer} IRS observations, the MIRI MRS observations were obtained early in the mission because the observations were part of a \emph{JWST} GTO program (PI Chen, Program ID 1294). Indeed, $\beta$ Pic was observed $\sim$6 months after the beginning of regular operations on January 11, 2023. For reference, \emph{JWST} launched on Christmas Day 2021 and began regular operations $\sim$6 months later in late June 2022. The observations were originally designed to both search for gradients in the properties of the debris disk as a function of stellocentric distance and characterize the mid-infrared spectrum of the $\beta$ Pic b companion. To facilitate comparison with the \emph{Spitzer} IRS observations, the \emph{JWST} MIRI observations were executed using all three grating settings (short, medium, long). Paper I describing the mid-infrared spectrum of $\beta$ Pic b has already been published \citep{worthen24}. This manuscript focuses on the properties of the dusty debris near the star. \footnote{The JWST data presented in this article were obtained from the Mikulski Archive for Space Telescopes (MAST) at the Space Telescope Science Institute. The specific observations analyzed can be accessed via \dataset[doi:10.17909/j6kp-2r12]{https://doi.org/10.17909/j6kp-2r12}}

To enable both debris disk and exoplanet science goals, the observational sequence contained four separate observations that were executed back-to-back using a non-interrupt (see Table~\ref{tab:mrs}). First, the telescope was slewed to an area of the sky, adjacent to $\beta$ Pic, with no known background sources, and a background sky observation was made. Second, the telescope was blind pointed at $\beta$ Pic and a mini-mosaick was executed, observing two fields, one centered on the host star and the other adjacent to the south-west. The original intention of this observation was to observe the two flanking fields adjacent to the host star; however, there was a mistake in planning. Third, the MIRI executed an on source Target Acquisition and observed the central host star. This observation was placed last because of concerns that the host star might saturate the detector and create persistence within the detector for subsequent disk observations. Since the mini-mosaick pointing on the host star was obtained without Target Acquisition, its spectrum can not be combined with the Target Acquisition data to minimize systematics. Finally, the telescope was slewed to N Car, an A0V star located 7.6$\arcdeg$ from $\beta$ Pic, to measure the PSF and enable classical PSF subtraction. The observation of the reference star required a Target Acquisition to ensure that the reference star would be placed at a similar location as the target in the instrument field-of-view. The non-interrupt minimized the changes in the thermal state of the telescope between observations. The observations were constrained to occur at a time in which the MRS slices would be orthogonal to the disk midplane to facilitate spatially resolved spectroscopy. The analysis in this manuscript focuses on the dedicated pointing centered on the host star using Target Acquisition and uses the background observation for empirical background subtraction and the calibration star observation to refine the flux calibration.

The raw $\beta$ Pic, N Car, and background observations were downloaded from the MAST archive and processed using JWST pipeline \citep{bushouse23} version 1.12.5 using CRDS (Calibrated Reference Data System) context “jwst 1193.pmap”. We ran the pipeline using the same settings as described in Paper I. Specifically, the pipeline consists of 3 stages: \texttt{Detector1}, \texttt{Spec2}, and \texttt{Spec3}. In \texttt{Detector1}, we changed the default three group rejection threshold from 6$\sigma$ to 100$\sigma$. In \texttt{Spec2}, in addition to the fringe flat correction which is applied by default, we also applied the residual fringe correction that is not applied by default. In \texttt{Spec2}, we performed the stray light subtraction that corrects the detector cross-artifact. In \texttt{Spec3}, we mosaicked the individual data cubes together using the \texttt{IFUalign} coordinate system so that the data cubes were aligned with the instrument field-of-view. Mosaicking in sky coordinates requires rotation and therefore interpolation of the data cubes. We elected to use \texttt{IFUalign} to minimize resampling. In \texttt{Spec3}, we also performed background subtraction using the dedicated background observations along with the outlier rejection step. Finally, we constructed the spectral cubes using the drizzle algorithm in the pipeline \citep{law23}, using the default spaxel size. The left portion of Figure~\ref{fig:Aperture_compare} shows the 15.4 $\mu$m slice of our data cube extracted from Channel 3C.

We extracted the spectrum of the unresolved point source in the $\beta$ Pic data cube at the location of the host star using aperture photometry. We refined the flux calibration of the spectrum using a Relative Spectral Response Function (RSRF) created from the N Car observation. We extracted two point source spectra using two different apertures: (1) a small aperture with a radius equal to 1.5$\times$FWHM of the unresolved JWST PSF for each slice and (2) a large aperture with radius equal to the angular resolution of the \emph{Spitzer} IRS, 11.5$\times$FWHM of the unresolved JWST PSF for each slice. Using the \emph{Spitzer} IRS extraction aperture, that covered the same region of the $\beta$ Pic disk and adjoining sky, was important to demonstrate robustly changes in the $\beta$ Pic spectrum between 2004/5 and 2023. An earlier version of the small aperture spectrum is presented in Paper I. We centered the apertures on the center of the unresolved PSF. To find the center, we collapsed each spectral cube along the wavelength axis and fit a 2-D Gaussian to the resulting image. Next, we applied a correction at the 1-D level to remove the 12.2 $\mu$m spectral leak \citep{gasman23}. Finally, we applied our N Car RSRF. The N Car 1-D spectrum was extracted using the same process as that used for the $\beta$ Pic 1-D spectrum described above. To create the RSRF, we used the stellar photosphere model from Paper I that fit a BT-NextGen stellar atmosphere model \citep{allard11} with $T_{*}$ = 8800 K, $\log g$ = 4.0, and stellar metallicity to the stellar ultra violet through near-infrared photometry. We empirically found that the RSRF not only improved the alignment of the sub-band spectra but also removed some systematic noise.

Once we extracted the calibrated $\beta$ Pic spectrum, we subtracted the stellar photosphere to reveal the thermal infrared excess spectrum from the circumstellar dust. We used the stellar photosphere model from \cite{lu22} that fit a Kurucz stellar atmosphere model with $T_{*}$ = 8000 K, $\log g$ = 4.0, and stellar metallicity to the stellar ultra violet through near-infrared photometry and spectra. Figure~\ref{fig:Aperture_compare} shows a comparison of the infrared excess spectra extracted using both the small (with diameter 1.5$\times$FWHM) and large apertures (with a diameter set by the \emph{Spitzer} angular resolution). As expected, the larger aperture includes more flux, consistent with the \emph{Spitzer} observations.

There are some minor features in our MRS spectrum that are probably artifacts. For example, the MRS contains a 6.1 $\mu$m light leak that manifests itself as additional 12.2 $\mu$m emission. The pipeline corrects for the light leak using observations of calibration stars. The correction assumes that the target is a point source. However, the $\beta$ Pic disk is extended throughout this wavelength region, making correction of the light leak more challenging than usual. Indeed, our Channel 3 A spectrum contains an incomplete correction of the 6.1 $\mu$m light leak which appears as a slight dip in the spectrum at 12.2 $\mu$m. In addition, A-type stars possess hydrogen absorption lines in their atmospheres: Humphreys-$\gamma$ (9$\rightarrow$6) at 5.908 $\mu$m, Pfund-$\alpha$ (6$\rightarrow$5) at 7.4599 $\mu$m, Humphreys-$\beta$ (8$\rightarrow$6) at 7.502 $\mu$m, and the $12\rightarrow$8 transition at 10.502 $\mu$m. The width and the depth of these lines vary from star to star depending on their stellar properties (such as $v\sin i$). We attemped to remove the stellar absorption features from our spectrum by spline fitting the observed N Car stellar atmosphere  features when creating our RSRF. However, our RSRF calibrator, N Car (A0II), is a slightly earlier type star than $\beta$ Pic (A6V); therefore, our correction is incomplete and leaves small residuals near the stellar absorption features.  

\section{Dust Modeling}

\begin{figure}
    \centering
    \includegraphics[scale=0.5]{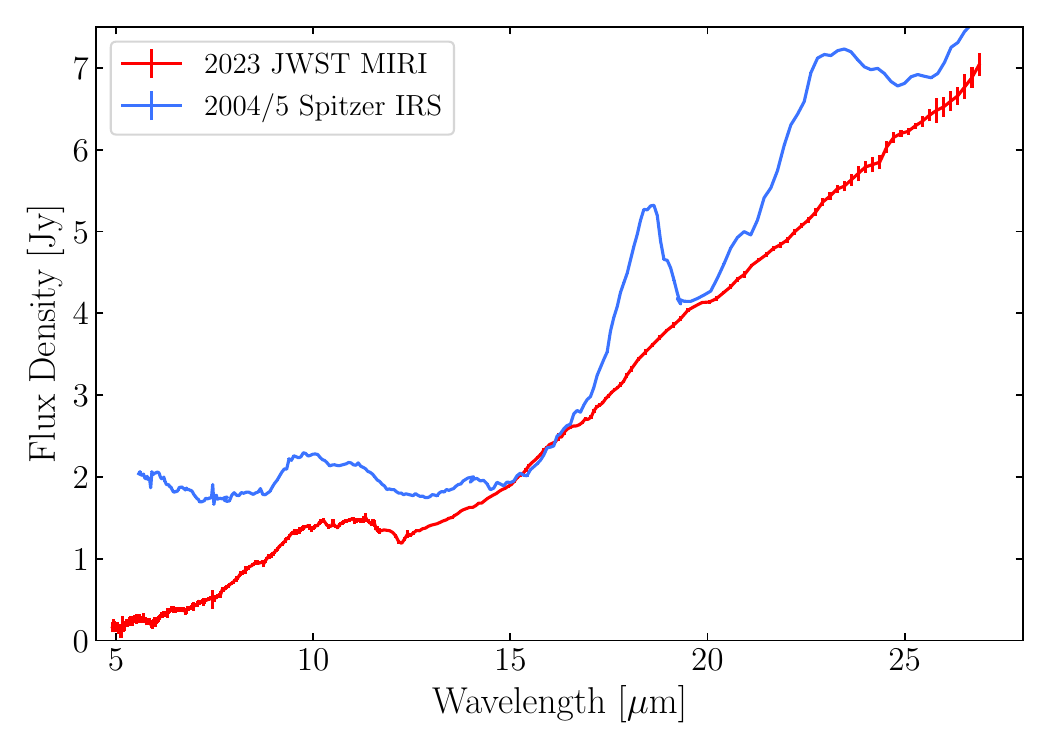}
    \caption{Comparison of the photosphere subtracted, mid-infrared spectra of $\beta$ Pic from the 2004/5 Spitzer IRS (blue) and 2023 \emph{JWST} MIRI observations (red). The MIRI MRS spectrum was extracted using an extraction aperture equal to the angular resolution of \emph{Spitzer}. Even with the Spitzer-sized aperture, the 18 $\text{\textmu}$m silicate feature seen with the \emph{Spitzer} IRS is not present in the \emph{JWST} MIRI spectrum.}
    \label{fig:Spitz_vs_MRS}
\end{figure}

We found that the $\beta$ Pic mid-infrared spectrum at 5 - 27 $\mu$m changed substantially between the 2004-5 \emph{Spitzer} IRS epoch and the 2023 \emph{JWST} MIRI MRS epoch (see Figure~\ref{fig:Spitz_vs_MRS}). We compared the two mid-infrared spectra using an equivalent extraction aperture to ensure that our MIRI spectrum covered the same physical area of the disk as our \emph{Spitzer} IRS extraction. Specifically, our large MIRI extraction aperture was designed to be equivalent to the one used for the \emph{Spitzer} IRS spectrum. Both observations have been background subtracted using empirical measurements of the sky and obervatory background. Since both spectra were measured with high SNR ($>$ 100), the differences in the continuum and silicate emission features are significant. In particular, we find three changes in the mid-infrared spectrum: (1) the infrared continuum at 5 - 15 $\mu$m has decreased, (2) the shape of the 10 $\mu$m silicate emission feature has changed, and (3) the 18 and 23 $\mu$m forsterite features have nearly disappeared. These changes are not expected to be the result in differences between how the spectra were extracted and calibrated. For example, the RSRFs adjusted the absolute flux calibration of individual orders or sub-bands by $<$20\% and removed systematic noise such as fringing at the few percent level.

To better understand how the dust grain population changed in the time between the two epochs, we performed detailed modeling of the new \emph{JWST} MIRI spectrum to compare with the already published models of the \emph{Spitzer} IRS spectrum. 

The first step in modeling the \emph{Spitzer} IRS spectrum was the removal of the majority of the 5 - 15 $\mu$m excess. This excess was modeled using a "hot" ($\sim$600 K) black body, that produced little to no excess at 3 $\mu$m and a 2 Jy excess at 5 $\mu$m \citep{lu22}. The dust mass in the hot component 
\begin{equation}
M_{d} = 16 \pi \tau \rho_s D_{cm}^2 a/3
\end{equation}
\citep{jura95} where $\tau$ is the dust optical depth, $\rho_s$ is the grain density, and $D_{cm}$ is the distance between the dust and star in centimeters. For the $\beta$ Pic hot dust, we estimate $\tau$ from the fractional infrared luminosity, $L_{IR}/L_*$ = 7.5$\times$10$^{-3}$, assuming a $\sim$600 K black body that produces a 2 Jy excess at 5 $\mu$m. If the dust is composed of silicates with $\rho_{s}$ = 3.5 g cm$^{-3}$ and $a$ = 1 $\mu$m and is located within a few au of the host star, similar to the distance of the hot dust discovered in the \emph{JWST} MIRI observations \citep{worthen24}, then $M_{dust}$ = 2.4$\times$10$^{23}$ g, approximately the mass of Vesta. This pre-treatment was not necessary for the \emph{JWST} MIRI spectrum because the hot excess was much diminished in the second epoch.

For the \emph{JWST} MIRI observation, we inferred the dust properties using only the photosphere subtracted spectrum. We were only able to model the mid-infrared spectrum over wavelengths 7.5 - 26.9 $\mu$m, a substantially shorter wavelength interval than the \emph{Spitzer} IRS observations. The MIRI MRS operates at wavelengths 5 - 30 $\mu$m, a smaller wavelength interval than the \emph{Spitzer} IRS low resolution modes (5.3 - 38 $\mu$m). We truncated the short wavelength end of the MRS spectrum at 7.5 $\mu$m because the fitting code attempted to reproduce wiggles in the spectrum at shorter wavelengths created by the incomplete correction of stellar absorption features. We truncated the long wavelength end of the spectrum at 26.9 $\mu$m because the spectrum was substantially noisier at longer wavelengths. Finally, we removed a small portion of the spectrum near 12.2 $\mu$m to ensure that our fits would not be impacted by the 6.1 $\mu$m light leak.

\begin{figure}
    \centering
    \includegraphics[scale=0.42]{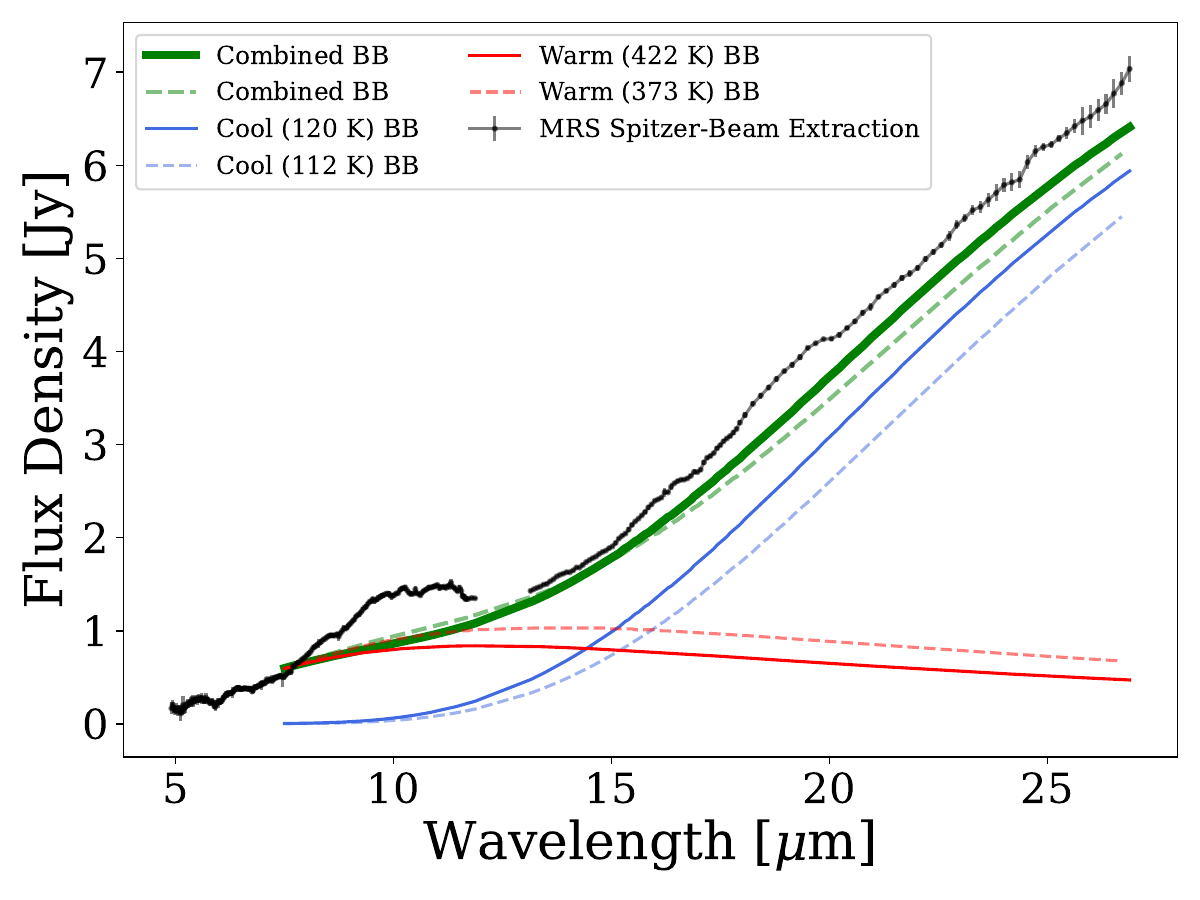}
    \caption{Black body continuum estimated from the 2023 MRS spectrum. The spectrum  extracted using the large aperture, consistent with that used for the Spitzer IRS observations, is plotted using black dots with error bars. Overlaid are the best fit warm (red line) and cool (blue line) black bodies and the sum of the two black body components (green line).}
    \label{fig:MRS_BB}
\end{figure}


We performed two fits of the MIRI MRS spectrum, one using the exact same set of optical constants used by \cite{lu22} for the 2004 IRS observations and another using slightly higher temperature forsterite and enstatite optical constants. Specifically, \cite{lu22} used "amorphous" olivine (Mg$_{2-2x}$Fe$_{2x}$SiO$_4$) and pyroxene (Mg$_{1-x}$Fe$_{x}$SiO$_3$) and their crystalline forms, forsterite and enstatite. Like \cite{lu22}, we assumed that the dust producing the solid-state silicate emission features had the same temperatures as the black bodies used to fit the continuum even though large black bodies grains are not co-spatial with small grains with the same temperature. We used optical constants for amorphous olivine with a chemical composition, MgFeSiO$_4$, and amorphous pyroxene, with a chemical composition, Mg$_{0.7}$Fe$_{0.3}$SiO$_3$ \citep{dorschner95}. We used optical constants for crystalline forsterite with a chemical composition, Mg$_{1.72}$Fe$_{0.21}$SiO$_4$, and crystalline enstatite with a chemical composition, Mg$_{0.92}$Fe$_{0.09}$SiO$_3$, measured at 10, 100, 200, 300, 551, 738, and 928 K \citep{zeidler15}. For the IRS equivalent fit, we used 100K and 300 K forsterite and enstatite optical constants. For the second fit, we used 100 K and 551 K forsterite and enstatite assuming that the warm dust population was hotter in the 2023 MIRI observation.

\subsection{Disk Continuum}

The dusty disk around $\beta$ Pic is believed to contain dust grains that radiate featureless continuum. A multi-wavelength analysis of high resolution scattered light and thermal emission images of $\beta$ Pic indicated that the bulk of the dust may be carbon-rich \citep{ballering16}. Carbonaceous grains lack solid state emission features at mid-infrared wavelengths. In addition, previous efforts to model the $\beta$ Pic mid-infrared spectrum used large black body grains to reproduce the continuum emission \citep{lu22, chen07}. To model the continuum in the MRS spectrum, we assumed that the cold dust population had a temperature, $T_c$, within the range (80, 160) K and the warm dust had a temperature, $T_w$, within the range (260, 800) K. To determine the best fit, we divided each temperature range into 7 steps, corresponding to 11 K and 77 K for the cold and warm dust populations, respectively. We found that the disk continuum was similar using either set of optical constants (100 K and 300 K forsterite or 100 K and 551 K forsterite). For the model including 300 K forsterite, we found a best-fit using a warm black body with a dust temperature, $T_{w}$ = 373 $\pm$ 80 K, and a colder black body with a dust temperature, $T_c$ = 112 $\pm$ 10 K. For the model including 551 K forsterite, we found a best-fit model using slightly higher temperature black bodies, $T_{w}$ = 422 $\pm$ 80 K and $T_c$ = 120 $\pm$ 10 K. In Figure \ref{fig:MRS_BB}, we overlaid the warm and cool black body components on our photosphere subtracted MIRI spectrum to show the contribution of the black body continuum to the overall spectrum.

\begin{figure}
    \centering
    \includegraphics[scale=0.4]{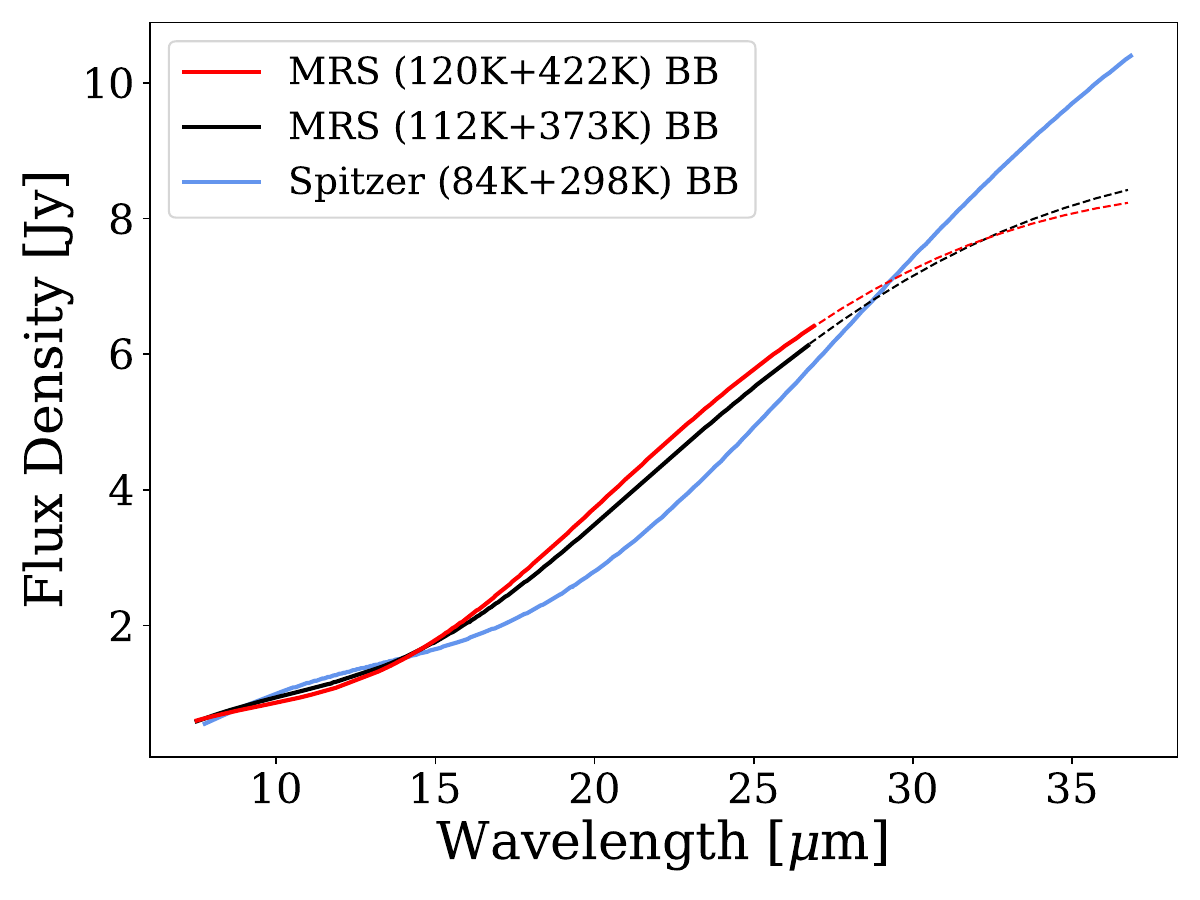}
    \caption{Comparison of the black body continuum from the 2004/5 Spitzer IRS (blue) and 2023 MRS MRS. The MIRI MRS observations have been fit using two different sets of forsterite optical constants. Eack fit uses a warm component and a cool $\sim$100 K component. The warm optical constants were measured at temperatures 300 K (black) and 551 K (red).}
    \label{fig:BBcomparisons}
\end{figure}

The warm and cold dust temperatures are broadly consistent with those found for the 2004 Spitzer IRS spectrum ($T_{w}$ = 298 $\pm$ 80 K, $T_{c}$ = 84 $\pm$ 10 K; \citep{lu22}). The uncertainties quoted in both fits represent the temperature fitting precision given the assumed dust grain temperature range and the number of temperature bins used for the fit, not the 1 $\sigma$ confidence level. Our fit captures the differences in the continuum by decreasing the dust mass needed for the warm black body component from 0.15 $M_{moon}$ to 0.03 - 0.05 $M_{moon}$ and shifting the temperature from 300 K to 373 - 422 K. It also decreases the dust mass needed for the cold black body component from 56 $M_{moon}$ to 12 - 14 $M_{moon}$ and shifts the temperature from 84 K to 112 - 120 K. In Figure \ref{fig:BBcomparisons}, we show the black body continuum fits for both the 2004 (after subtraction of the hot $\sim$600 K dust component) and 2023 epochs for comparison. Since the black body function used to model the continuum is much broader in wavelength than the silicate emission features, uncertainties in the continuum are not likely to impact the dust species identification that is made based on peak position. However, uncertainties in the continuum may impact the overall flux assigned to silicate emission features and therefore estimates of the dust mass.

\begin{figure}
    \centering
    \includegraphics[scale=0.4]{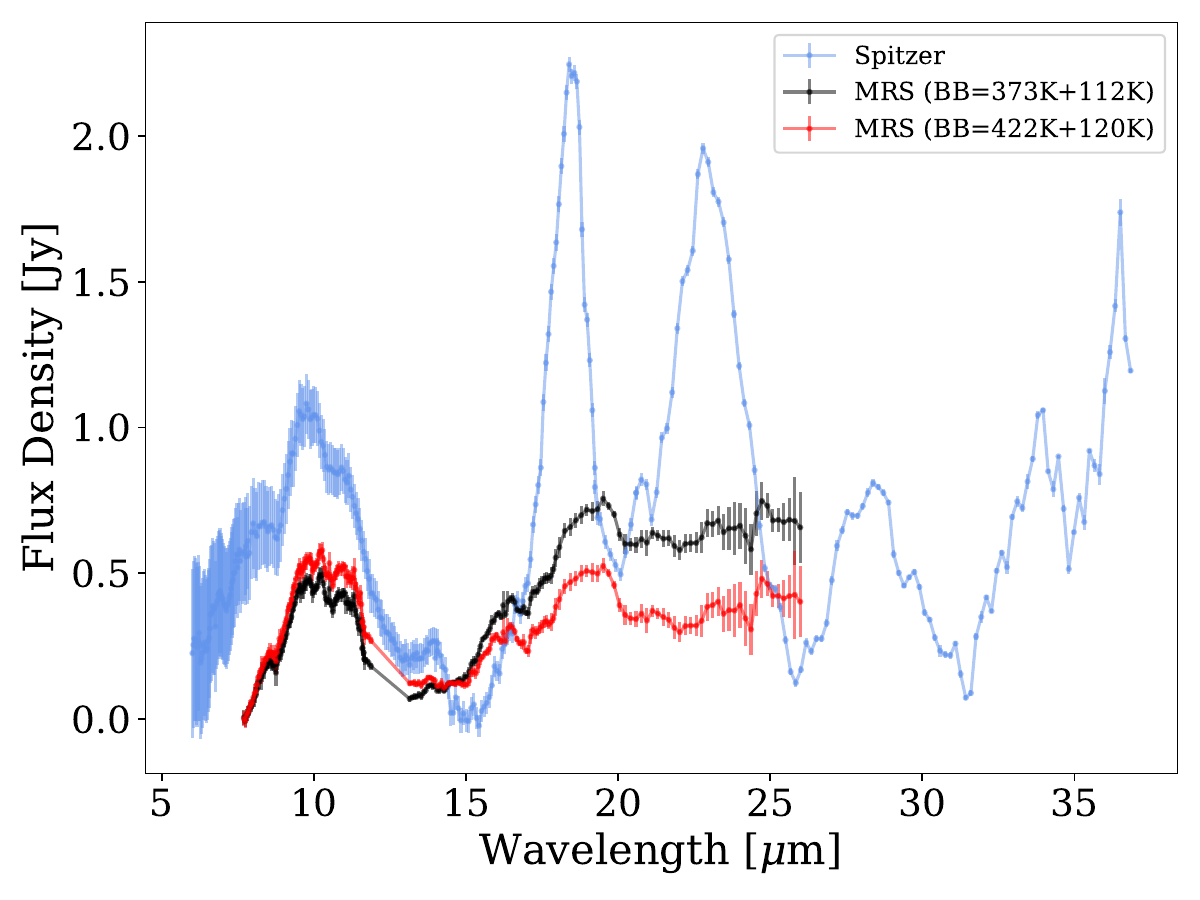}
    \caption{Comparison of the black body continuum subtracted, excess spectra from 2004/5 Spitzer IRS (blue) and MIRI MRS. The MIRI MRS fits using 300 K (black) and 551 K (red) forsterite optical constants are shown separately. This highlights the changes in the 10, 18, and 23 $\mu$m silicate emission features between the two epochs. The 10 $\mu$m feature has decreased in amplitude and the 18 and 23 $\mu$m silicate emission features, generated by cold crystalline silicates, have disappeared.}
    \label{fig:MRSvsIRS}
\end{figure}

\begin{figure*}[t!]
    \centering
    \includegraphics[scale=0.8]{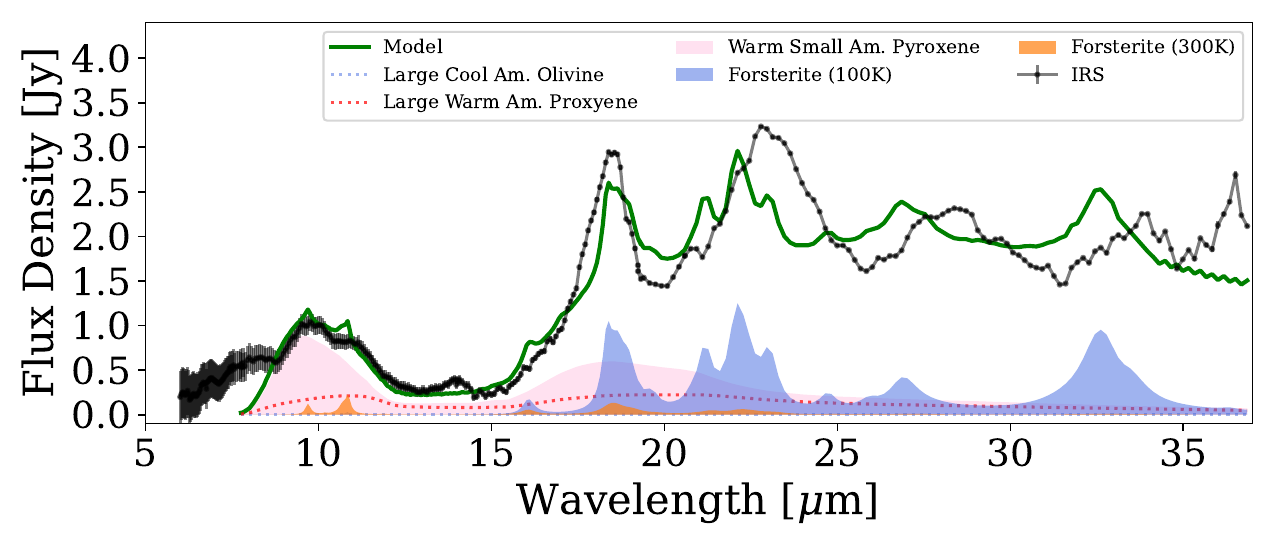}
    \caption{Black body subtracted 2004 \emph{Spitzer} IRS spectrum of $\beta$ Pic. Overlaid is the \cite{lu22} fit optimized to reproduce the overall shape of the 18 $\mu$m feature ($\chi^2$ = 23.6), using spherical particles and forsterite and enstatite optical constants measured at 100 K and 300 K. The fit reproduces the overall shape of the 10 $\mu$m feature well; although, it overpredicts the emission from warm crystalline forsterite and does not account for the short wavelength "shoulder" to the 10 $\mu$m feature that is emitted by an unidentified carrier. This 8.7 $\mu$m feature has been observed toward TW Hya and FN Tau and is probably generated by a polymorph of silica \citep{sargent06}. The fit does not reproduce the 23 and 28 $\mu$m features well probably because the disk is spatially extended and the two temperature model is not accurate. See \cite{lu22} for additional fits of the silicate features.}
    \label{fig:Model-04}
\end{figure*}

Our analysis is focused on reproducing the \emph{Spitzer} IRS and \emph{JWST} MIRI spectra and not reproducing the far-infrared to millimeter spectral energy distribution. In isolation, mid-infrared spectra provide limited constraints on the temperature and mass of the warm and cold dust continuum. Using the Wien Displacement Law, we estimate that the peaks of 120 K and 80 K black bodies occur at 24 and 36 $\mu$m, respectively. These wavelengths are similar to the long wavelength cut-offs for the MRS ($\sim$27 $\mu$m) and IRS ($\sim$37 $\mu$m). Thus, the simple two component black body model may either under or overpredict the dust thermal emission outside of the instrument spectral range. Indeed, Figure \ref{fig:BBcomparisons} shows that the black body continuum model for the \emph{JWST} MIRI spectrum is probably too low at $\lambda$ $>$ 26 $\mu$m because it is significantly lower than the \emph{Spitzer} IRS continuum and therefore underpredicting the mass of cold dust. Moreover, the disk temperature structure is expected to be more complicated because $\beta$ Pic has a large radial extent. As a result, we only use the two black body model as an approximate guide to (A) the temperature of the dust components and (B) the continuum emission so that detailed solid state feature analysis can be performed. We do not use the masses that can be estimated from the warm and cold black body fits in our analysis.

\subsection{Spectral Features}

Detailed fitting of the 2004 Spitzer IRS spectrum of $\beta$ Pic indicated that (1) warm dust with temperatures $\sim$300 - 500 K gave rise to both the amorphous and crystalline components of the 10 $\mu$m silicate emission, (2) the same warm dust also emitted a broad 20 $\mu$m amorphous silicate emission feature, and (3) cool dust with temperatures $\sim$100 K emitted the 18, 23, 28, and 33 $\mu$m crystalline silicate features \citep{lu22}. In Figure \ref{fig:MRSvsIRS}, we show the black body subtracted excess spectra for the 2004 and 2023 epochs to illustrate more clearly the changes in the silicate emission features. The 10 $\mu$m silicate feature appears to have both decreased in amplitude and changed in shape. The 18 and 23 $\mu$m crystalline silicate features appeared to have disappeared entirely. Since the 18 and 23 $\mu$m forsterite features were principally emitted by $\sim$100 K dust, a cursory examination of the spectrum indicates that this component has vanished in the 20 years since the \emph{Spitzer} IRS observation.

\subsubsection{2004 Epoch Fitting}

The IRS spectrum showed clear solid-state emission features at 10, 18, 23, 28, and 33 $\mu$m \citep{lu22, chen07}. The two component dust model fit by \cite{lu22} assumed that the two silicate populations had the same temperatures as the black bodies. For consistency, they used the 100 K and 300 K forsterite and enstatite optical constants for the cold and warm dust, respectively. While they were unable to fit the peak wavelengths and shapes for all of the silicate features simultaneously using a single model, they were able to find models that fit individual features well. For example, the 18 $\mu$m silicate emission feature was best fit using spherical grains while the 23 $\mu$m feature was best fit using grains with shapes well described by the Common Distribution of Ellipsoids 2 (CDE2) and the 28 and 33 $\mu$m features were best fit using grains with shapes well described by the Common Distribution of Ellipsoids 1 (CDE1), indicating a possible gradient in the shape of the dust grains from regular to irregular as a function of stellocentric distance. 

The \cite{lu22} model fit the overall short and long wavelength edges of the 2004 10 $\mu$m silicate feature well (see Figure \ref{fig:Model-04}). The 2004 10 $\mu$m feature appeared to be dominated by amorphous olivine with a smaller amount of crystalline forsterite. The \cite{lu22} model shown in Figure \ref{fig:Model-04} was optimized to reproduce the wavelength and the shape of the narrow, high line-to-continuum ratio 18 $\mu$m feature. The model slightly overpredicted the amount of warm forsterite emission and did not reproduce emission observed on the short wavelength side of the silicate emission feature at $<$8.7 $\mu$m. This "shoulder" has also been observed in the \emph{Spitzer} IRS spectra of the TW Hya and FN Tau protoplanetary disks; although, the carrier has not yet been definitively identified. \cite{sargent06} speculated that this short wavelength emission may indicate the presence of a high temperature polymorph of silica. Indeed, for the 2004 epoch, the IRS spectrum was fit over the wavelength range 7.7 - 36.8 $\mu$m to minimize the influence of the shoulder on the fit of the 10 $\mu$m feature.

\subsubsection{2023 Epoch Fitting}

\begin{figure*}[t!]
    \centering
    \includegraphics[scale=0.75]{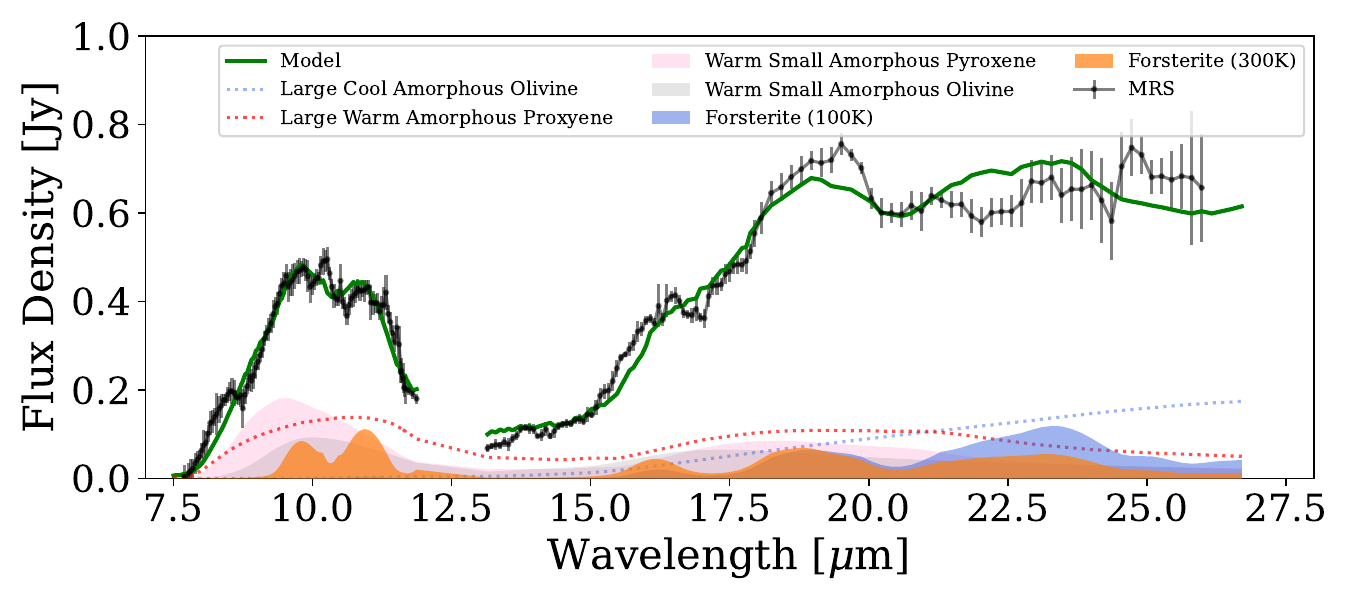}
    \caption{Black body subtracted 2023 MIRI MRS spectrum of $\beta$ Pic. Overlaid is the best fit model using forsterite and enstatite optical constants measured at 100 K and 300 K ($\chi^2$ = 2.0), the same optical constants used to fit the \emph{Spitzer} IRS spectrum. This fit has slight differences in the short wavelength rise and the long wavelength decline in the 10 $\mu$m silicate feature. The long wavelength side of the feature declines at slightly longer wavelengths than predicted using the 300 K optical constants. The short wavelength side of the 10 $\mu$m feature is not well fit because the carrier responsible for the "shoulder" still has not yet been identified. The error bars for the MRS spectrum grow large enough beyond 21 $\mu$m that the model fit is not too discrepant with the data.}
    \label{fig:Model-300K}
\end{figure*}

We initially fit our MRS spectrum using 100 K and 300 K forsterite and enstatite optical constants, consistent with the 2004 model. We found that using the same optical constants provided a reasonable fit to the 2023 10 $\mu$m silicate feature than the 2004 10 $\mu$m silicate feature. Looking more closely, we found that our best fit model did not perfectly reproduce either the short-wavelength rise in the 10 $\mu$m feature or the long wavelength decline. We plotted the black body subtracted excess spectrum from the 2023 MRS observations with the silicate emission model using 100 K and 300 K forsterite and enstatite optical constants overlaid in Figure \ref{fig:Model-300K}. At short wavelengths, we struggled to fit the silicate feature because the unidentified shoulder still could not be fit with a known dust species. At long wavelengths, we discovered that the 10 $\mu$m silicate feature was broader than predicted by our 300 K forsterite optical constants. 

\begin{figure*}[t!]
    \centering
    \includegraphics[scale=0.77]{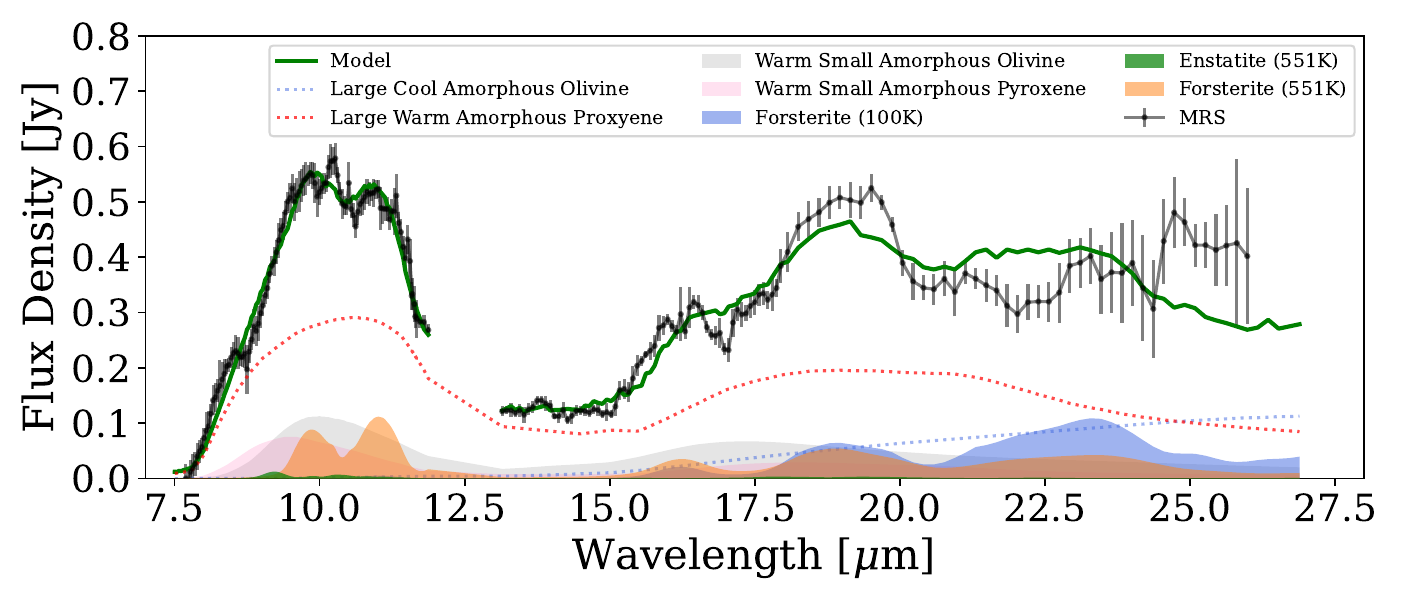}
    \caption{Same as above. Overlaid is the best fit model ($\chi^2$ = 1.65) using forsterite and enstatite optical constants measured at 100 K and 551 K. This fit reproduces the long wavelength decline of the 10 $\mu$m feature better, suggesting that the warm dust component may have increased in temperature since 2004.}
    \label{fig:Model-551K}
\end{figure*}

Since laboratory measurements indicate that solid-state features shift to longer wavelengths with increasing temperatures, we refit the spectrum using 100 K and 551 K forsterite and enstatite optical constants. We chose the 551 K optical constants even though 551 K is somewhat hotter than our best fit warm black body temperature because 551 K was the next warmest temperature at which the optical constants were measured. We plotted the black body subtracted excess spectrum from the 2023 MRS observations with the silicate emission model using 100 K and 551 K forsterite and enstatite optical constants overlaid in Figure \ref{fig:Model-551K}. In this case, we found that the amorphous olivine optical constants, combined with the 551 K Forsterite optical constants, better reproduced the long wavelength decrease. However, our fit still struggled to reproduce the short-wavelength rise in the 10 $\mu$m feature without an additional dust component to describe the short wavelength shoulder. Our fits therefore suggest that dust producing the 2023 10 $\mu$m silicate emission may be warmer than the dust that produced the 2004 10 $\mu$m silicate emission.

\begin{deluxetable*}{lcccc}
\tablecaption{Dust Masses for the Best-Fit Models of the $\beta$ Pic debris disk}\label{tab:chi2models}
\tablehead{
\colhead{Species} & \colhead{Shape} & \colhead{2004 IRS} & \colhead{2023 MRS (100/300 K)} & \colhead{2023 MRS (100/551 K)} \\
\colhead{} & \colhead{} & \colhead{(M$_{moon}$)} & \colhead{(M$_{moon}$)} & \colhead{(M$_{moon}$)}}
\startdata
\multicolumn{2}{l}{Hot Dust Continuum Temperature (T$_{\textbf{h}}$ )}                & $\sim$600 K & N/A & N/A \\
\hline
Black body & \nodata & $3.2 \times 10^{-3}$ & N/A & N/A  \\
\hline
\multicolumn{2}{l}{Warm Dust Continuum Temperature (T$_{\textbf{w}}$)}                & $298 \pm 11$\,K & $373 \pm 70$\,K  & $422 \pm 70$\,K \\
\hline
Black body & \nodata & $(1.46 \pm 0.06) \times 10^{-1}$ & $(5.0 \pm 0.12) \times 10^{-2}$ & $(2.8 \pm 0.08) \times 10^{-2}$ \\
Pyroxene (Mg$_{0.7}$Fe$_{0.3}$SiO$_{3}$)         &CDE2, Rayleigh Limit                   & $(4.6 \pm 0.6) \times 10^{-5}$ & $(3 \pm 0.8) \times 10^{-6}$ & $(8 \pm 5) \times 10^{-7}$ \\
Pyroxene       &Mie, $5\,\mu m$ radius, $60$\% porosity  & ($1.7 \pm 0.6) \times 10^{-5}$ & $(5 \pm 1) \times 10^{-6}$ & $(7 \pm 0.8) \times 10^{-6}$ \\
Olivine (MgFeSiO$_{4}$)         &CDE2, Rayleigh Limit                   & $(0 \pm 4) \times 10^{-6}$ & $(1.9 \pm 0.8) \times 10^{-6}$ & $1.5 \pm 0.5) \times 10^{-6}$ \\
Olivine        &Mie, $5\,\mu m$ radius, $60$\% porosity  & $(0 \pm 5) \times 10^{-6}$ & $(0 \pm 1) \times 10^{-6}$ & $0 \pm 8) \times 10^{-7}$   \\
Forsterite (Mg$_{1.72}$Fe$_{0.21}$SiO$_{4}$)                      & (1)                   & $(8.2\pm 8.1) \times 10^{-8}$ & ($3.4\pm 1.9) \times 10^{-8}$  & ($2.3 \pm 1.3) \times 10^{-8}$ \\
Enstatite  (Mg$_{0.92}$Fe$_{0.09}$SiO$_{3}$)      & (1)                   & $(0 \pm 8)\times 10^{-8}$ & $(0 \pm 2)\times 10^{-8}$ & $(0.1 \pm 1.0)\times 10^{-8}$ \\
\hline
\multicolumn{2}{l}{Cool Dust Continuum Temperature (T$_{\textbf{c}}$ )}                & $84 \pm 7$\,K & $112\pm 11$\,K & $120\pm 11$\,K \\
\hline
Black body & \nodata & $(5.6 \pm 0.02) \times 10^1$ & $(1.4 \pm 0.04) \times 10^1$ &  $(1.2\pm 0.03) \times 10^{1}$  \\
Pyroxene (Mg$_{0.7}$Fe$_{0.3}$SiO$_{3}$)       &CDE2, Rayleigh Limit                  & $(0 \pm 2.8) \times 10^{-3}$ & $(0 \pm 4) \times 10^{-4}$  & $(0 \pm 3) \times 10^{-4}$ \\
Pyroxene      &Mie, $5\,\mu m$ radius, $60$\% porosity & $(0\pm 2.2) \times 10^{-3}$ & $(0 \pm 4) \times 10^{-4}$ & $(0\pm 3 \times 10^{-4}$ \\
Olivine (MgFeSiO$_{4}$)        &CDE2, Rayleigh Limit                  & $(3.45 \pm 0.23) \times 10^{-2}$ & $(8\pm 3) \times 10^{-4}$ & $(0 \pm 2) \times 10^{-4}$ \\
Olivine       &Mie, $5\,\mu m$ radius, $60$\% porosity & ($0\pm 1.4) \times 10^{-3}$ & $(4 \pm 3) \times 10^{-4}$ & $(2 \pm 2) \times 10^{-4}$  \\
Forsterite (Mg$_{1.72}$Fe$_{0.21}$SiO$_{4}$)                     & (1)                  & $(2.514\pm .507)\times 10^{-4}$ & $(4.1\pm 6.3)\times 10^{-6}$ & $(2.7\pm 4.4)\times 10^{-6}$ \\
Enstatite (Mg$_{0.92}$Fe$_{0.09}$SiO$_{3}$)                    & (1)                  & $(1.83\pm 5.64) \times 10^{-5}$ & $(1.4\pm 8.6) \times 10^{-6}$& ($2\pm 58)\times 10^{-7}$  \\
\hline 
$\chi^2$                       & \nodata                &$23.6$ & $2$ & $1.65$\\
\enddata
\tablecomments{(1) The shape distribution for forsterite and enstatite are CDE1 for 10 $\mu$m feature, Mie for IRS Spectral fit focusing on the 18 $\mu$m feature, CDE2 for all MRS fits. (2) The warm dust species share the same stoichiometry as the cold dust species. (3) The optical constants for the forsterite and enstatite used for the cool dust are measured at $100$ K, while those for the warm dust are measured at either 300 K or $551$ K. Therefore, their mass fraction coefficients are independent from each other. (4) The $\chi^2$ values calculated for the best-fit models use the full wavelength range of each spectra respectively ($7.5$--$26~\mu$m for the MRS spectrum, 7.7–36.8 $\mu$m for the IRS spectrum).}
\end{deluxetable*}

In general, both fits predict the amplitudes of the 16 and 18 $\mu$m forsterite features well but overpredict the emission expected from the 23 $\mu$m crystalline forsterite feature. One possibility is that our model is too simplistic. To simplify the modeling, we assumed that all of the dust could be described using two temperatures. If all of the dust at one temperature is located at one distance, then the model suggests that the dust is located in two relatively narrow rings. Multiwavelength imaging of the dust around $\beta$ Pic indicates that the dust is not located in two distinct planetesimal belts but is located in an extended disk with an inner radius ($\sim$ few au) and an outer radius. High resolution mid infrared imaging indicates that micron sized grains are located at distances up to 300 au \citep{worthen24,rebollido24,telesco05} and ALMA observations indicate that the millimeter-sized grains are located at distances up to 120 au \citep{matra19,dent14}. Another possibility is that the Iron:Magnesium ratio of the silicates could be different than we assumed. Pure forsterite (Mg$_{2}$SiO$_{4}$) has a slightly more precipitous decline between its 18 and 23 $\mu$m features than the Fo90 forsterite used in our fits \citep{koike03,chihara02}.

Although no single fit captured the 10 $\mu$m feature well, we were able to draw general conclusions about the changes in the dust emitting the solid-state silicate features (see Table~\ref{tab:chi2models}): (1) The mass of warm, amorphous silicates (olivine and pyroxene) decreased from $\sim$6$\times$10$^{-5}$ $M_{moon}$ to $\sim$1$\times$10$^{-5}$ $M_{moon}$ and the size of these dust grains shifted from predominantly small grains in the Rayleigh Limit to include more large, porous 5 $\mu$m grains that could be described using Mie Theory. (2) The mass of small, warm crystalline silicates (forsterite and enstatite) in the Rayleigh Limit decreased from 8.2$\times$10$^{-8}$ $M_{moon}$ to 2.4-3.4$\times$10$^{-8}$ $M_{moon}$. (3) The mass of cool $\sim$100 K amorphous silicates (olivine and pyroxene) decreased from 3.45$\times$10$^{-2}$ $M_{moon}$ to 0.2-1.2$\times$10$^{-4}$ $M_{moon}$ and the size of these dust grains shifted from predominantly small grains in the Rayleigh Limit to include more large, porous 5 $\mu$m grains that could be described using Mie Theory. (4) Finally, the mass of small, cool $\sim$100 K crystalline silicates (forsterite and enstatite) in the Rayleigh Limit decreased by a factor of 100 from 2.7$\times$10$^{-4}$ $M_{moon}$ to 0.29-5.5$\times$10$^{-6}$ $M_{moon}$. It is this final component that gave rise to the prominent 18, 23, 28, and 33 $\mu$m forsterite features observed in the 2004 IRS data. Thus, the absence of these features in the MRS data signals the absence of this dust component in the $\beta$ Pic disk in 2023.

\section{Discussion}

\subsection{The Disappearance of the hot continuum and the cold crystalline features}

The two most striking changes in the $\beta$ Pic mid-infrared spectrum between 2004-5 and 2023 are (1) the disappearance of the $\sim$600 K hot dust continuum and (2) the disappearance of the 18 and 23 $\mu$m crystalline forsterite features. One possible explanation for their disappearances is that the grains creating both signatures are small and that radiation pressure has blown the small dust grains away. Indeed, the spectral features in the \emph{Spitzer} IRS spectra were modeled assuming dust grains in the Rayleigh Limit (2$\pi a$ $\ll$ $\lambda$). In this case, a large collision would have created the fine dust particles prior to the \emph{Spitzer} IRS observations. Radiation pressure would have acted on the grains driving them out radially so that they had particular distances and temperature in 2004-5 when the system was observed by \emph{Spitzer} IRS. If there was one giant collision, then the particles that created the 18 and 23 $\mu$m features could have been composed of the smallest particles that were able to travel the furthest. The particles that created the hot dust component could have been composed of larger grains that were not able to travel as far. In this scenario, the grains would have continued on their outward journey after 2004-5. By 2023, they could have been either so cold that they no longer radiated heat in the MIRI MRS passband or at such large distances that they were no longer within the MIRI MRS field of view.

For the circumstellar environment around an A-type star such as $\beta$ Pic, the dynamics of a dust grain is determined by the ratio of the force due to radiation pressure compared with that due to the gravity acting on the grain, which is given by the parameter $\beta$, where 
\begin{equation}
    \beta=\frac{3L_*\langle Q_{pr}\rangle}{16\pi GM_*ca\rho}.
\end{equation}
\citep{artymowicz88} where $L_*$ is the stellar luminosity, $M_*$ is the mass of the star, $c$ is the speed of light, $G$ is the gravitational constant, $a$ is the radius of the dust grains, $\rho$ is the bulk density of the material, and $\langle Q_{pr}\rangle$ is the average radiation pressure coupling coefficient for a given material. Grains with $\beta$ $>$ 0.5 are expected to be blown out of the system. For $\beta$ Pic with $L_*$ = 8.7 $L_{sun}$ and $M_{*}$ = 1.2 $M_{sun}$, silicate grains with densities $\rho$ = 3.5 g cm$^{-3}$ and sizes $a$ $<$ 2 $\mu$m are expected to be radiatively driven from the system. For example, grains with $a$ = 0.5 and 1.0 $\mu$m are expected to have $\beta$ = 1 - 2 and 0.5 - 1, respectively, depending on their composition \citep{worthen24}. Since the grains used to model the 18 $\mu$m feature are in the Rayleigh Limit (2$\pi a \ll \lambda$), they are expected to have a radius, $a\ll$ $\lambda/2\pi$ = 18 $\mu$m/2$\pi$ = $2.9$ $\text{\textmu}$m. As a result, they are expected to be sensitive to radiation pressure. The same is true for the dust grains responsible for the hot dust component. 
 
 We estimate the blowout timescales of the silicate grains that produced the 18 $\text{\textmu}$m feature seen in the \emph{Spitzer} IRS spectrum by first estimating their terminal velocity. The terminal velocity of a dust grain under the influence of radiation pressure and gravity is
 \begin{equation}\label{eq:term_v}
     v_r\simeq [(2GM_*/r_{init})(\beta-1/2)]^{1/2}
 \end{equation}
 \citep{su09} where $r_{init}$ is the initial stellocentric distance for the dust grains. We calculate the terminal velocity and the distance a dust grain traveled between the \emph{Spitzer} and \emph{JWST} observations for different sized grains. If the silicates producing the 18 $\text{\textmu}$m feature have a radius of 0.5 $\mu$m and started at the minimum distance for their temperature of 9 au \citep{lu22}, we calculate a terminal radial velocity of the dust grains of 23 km/s if the grains have a composition of Mg$_{1.72}$Fe$_{0.21}$SiO$_4$, consistent with the \cite{lu22} fits. If the 0.5 $\text{\textmu}$m grains traveled at this velocity for the entire 19 years between observations with \emph{Spitzer} and \emph{JWST}, they would have traveled a total radial distance of $\sim$90 au, which would put them just ouside the field-of-view of the MRS observations in Channel 3. 
 
 We repeat the same calculation but for larger grains with radii of 1 $\text{\textmu}$m, that are also representative of the population giving rise to the hot dust component. For grains with a radius of 1 $\text{\textmu}$m, the terminal velocity is 10 km/s. The distance that 1 $\text{\textmu}$m sized grains would travel at this velocity between \emph{Spitzer} and \emph{JWST} observations is 41 au. This would put them at a distance of 50 au at the time of the \emph{JWST} observations. At a distance of 50 au from $\beta$ Pic, the blackbody radiative equilibrium temperature of the dust is $\sim$ 65 K, which has a peak blackbody wavelength of $\sim$45 $\text{\textmu}$m. Grains at this temperature would not emit significant flux at 18 $\text{\textmu}$m which could explain why we do not see the 18 $\text{\textmu}$m silicate feature in the MRS spectrum if micron sized grains produced the 18 $\text{\textmu}$m feature seen with \emph{Spitzer}. The same is true for the micron-size grains initially located at a few au responsible for the host dust component observed in the 2004 epoch: by the time of the second epoch, with a terminal velocity of $\sim$ 10 km/s, these grains would have a much diminished contribution to the 5 - 15 $\mu$m excess emission. 
 
 One explanation for the changes in the silicate features and the hot dust component between different observing epochs \emph{Spitzer} IRS and \emph{JWST} MIRI is that the silicate grains that are detected around $\beta$ Pic are not produced in a steady state manner over the 18-19 years between observations, but rather stochastically. The observed silicate features that are produced by small grains, like those seen with \emph{Spitzer}, would then be products of recent collisions between planetesimals and not collisional processes that are in steady state over
 timescales much longer than a decade. Then, over the time between observations, the silicate grains are driven outwards by radiation pressure and are in a region of the system where they are no longer detectable in our MRS observations. 

As we mentioned earlier, there are other young debris disks systems (35—80 Myr), the so called extreme debris disks (dustier than $\beta$-Pic, with fractional infrared luminosities of $L_{IR}/L_{*}$ $>$ 0.01), that have also been found to have very significant variability in the inner disk (terrestrial planet) region, on spatial scales similar to the location of the hot-dust component in $\beta$-Pic. Some of these systems showed variability both short-term (weekly to monthly) and long-term (yearly) \citep{su19}, while others were found to be quiescent for 3 years before showing large changes in dust production \citep{su22, rieke21}. Because the IRS and MIRI data are separated by 18-19 years and there are no mid-infrared observations in the intervening period, it is not possible to know the “duty cycle” of the event(s) giving rise to the hot dust component and the 18 $\mu$m and 23 $\mu$m crystalline features in the first epoch, pointing to the need of long-term monitoring. Even if there had been no dust production events in the intervening period, this variable behavior on a few decades timescale could be of the same nature as that of the extreme debris disks, just less frequent and leading to lower levels of dust production. To place these stochastic collisional events in a broader context, below we discuss other evidence of recent large collisions in the $\beta$-Pic system.  

\subsection{Giant collisions in the beta Pic disk}

Recent \emph{JWST} MIRI coronagraphic observations have revealed the presence of a narrow filament of dust named the cat's tail. Dynamical models are consistent with a collisional origin for the cat's tail in which a giant collision occurred near the present-day base of the cat's tail, $\sim$85 au from the host star, $\sim$150 years ago \citep{rebollido24}. During the time since the collision, radiation pressure blew unbound grains out of the exoplanetary system until they entered the interstellar medium and were no longer strongly influenced by the local, stellar environment. In their model, \cite{rebollido24} find that grains with a $\beta$ = 0.5 - 10 could have created the cat's tail if the direction of the collision was approximately parallel to the observer's line-of-sight to the star. This alignment is necessary to give the cat's tail a relatively short angular appearance on the sky ($\sim$100 au) despite having a much larger physical size ($\sim$900 au). In this case, radiation pressure sorts the grains so that the smallest ones travel the furthest and the largest ones remain closest to the star. They further estimate a 65\% probability of at least one collision of a 100 km or larger body (in radius) and a 0.5\% probability of a collision of a 500 km or larger body during the past 150 years though they note that these estimates are highly uncertain.

The collision captured by the \emph{Spitzer} IRS is probably not physically related to the cat's tail or any giant collisions that may have created the cat's tail. Although the \emph{Spitzer} IRS collision was not imaged, our modeling suggests that it likely occurred much closer to the star than 85 au and much more recently than 150 years ago. We can estimate the distance at which the \emph{Spitzer} IRS collision occurred assuming that the collision occurred at the location of the hot dust in 2004-5. This sets an upper limit on the distance of the collision from the host star. If the 600 K dust is comprised of black bodies, then they were located at 0.7 au from the host star in 2004-5. If it was comprised of small grains (2$\pi a$ $\ll$ $\lambda$), then they were located at 2.3 au from the host star in 2004-5. Both 0.7 au and 2.3 au are substantially closer to the host star than 85 au. Similarly, the \emph{Spitzer} IRS collision must have occurred much more recently than 150 years ago. If the \emph{Spitzer} IRS dust had been created 150 years ago, then the dust would have been removed by radiation pressure long before 2004-5 because the particles are very small and can be well modeled in the Rayleigh Limit (2$\pi a$ $\ll$ $\lambda$). Given the rapid timescale on which sub-micron sized grains are removed from the terrestrial planet zone, we hypothesize that the collision that created the dust probably occurred within the decade just before the 2004-5 observations.

New \emph{JWST} observations are needed to understand the frequency with which giant collisions occur in the terrestrial planet zone around $\beta$ Pic and whether these collisions are related to the cat's tail. So far, giant collisions within the terrestrial planet zone have been most easily detected by monitoring the mid-infrared spectrum; however, $\beta$ Pic has only been comprehensively observed using space-based mid-infrared spectroscopy at two epochs separated by $\sim$20 years. New MIRI MRS observations are needed to determine how often $\beta$ Pic experiences giant collisions in it's terrestrial planet zone. In addition, MIRI Imaging observations are needed to definitively determine whether the mid-infrared variability is related to the cat's tail. Unfortunately, the already existing MIRI Coronagraphic observations have an inner working angle of 3 $\arcsec$, corresponding to 60 au. Therefore, the terrestrial planet zone is obscured by the coronagraph. MIRI Imaging observations have a much smaller inner working angle ($<$1 $\arcsec$) that can provide access to regions closer to the terrestrial planet zone. As a result, future MIRI Imaging observations could be used to search for dust structures that directly connect the cat's tail with the terrestrial planet zone.

\subsection{Consequences for Planetary Mass Companions}

The existence of an outflowing wind of sub-blowout sized grains has been previously inferred for the low activity, steady-state circumstellar environment around $\beta$ Pic. In Paper I, \cite{worthen24} show that the debris disk is spatially resolved at 5 $\mu$m and that the grain temperature inferred from the \emph{JWST} MRS observations ($\sim$500 K) is consistent with the presence of small grains in the Rayleigh Limit. They show that if the grains are created in collisions at 0.9 au, the black body distance for 500 K dust, then the accretion rate onto $\beta$ Pic c and b is expected to be $\sim$2-3$\times$10$^{-15}$ $M_{Jup}$ yr$^{-1}$ and $\sim$2-3$\times$10$^{-17}$ $M_{Jup}$ yr$^{-1}$, respectively, significantly less than the accretion rate expected onto protoplanetary disk companions PDS 70 c and b, $\sim$10$^{-7}$ M$_{Jup}$ yr$^{-1}$ \citep{wang20}. Comparison of the accretion rates of protoplanetary and debris disks must be done carefully because the debris disk accretion rate only includes dust while that of protoplanetary disks typically includes both gas and dust even though the gas may not be directly detected. As a result, only the dust accretion rate onto planets in protoplanetary disks should be compared with that onto planets in debris disks. Therefore, the PDS accretion rates should be divided by 100, the canonical gas:dust ratio assumed in protoplanetary disks.

Our comparison of the 2004-5 \emph{Spitzer} IRS observations with the 2023 \emph{JWST} MIRI observations shows that a giant collision created a large mass of fine crystalline forsterite dust that was radiatively driven from the system. The dispersal of this material during the $\sim$20 year period enables us to make a better estimate for accretion rates onto the c and b components under the same assumptions, namely that the dust originated from 0.9 au, the same region from which the current outflowing wind is hypothesized to originate. The planetary mass accretion rate is 
\begin{equation}\label{eq:Mdot}
\dot{M}=\frac{M_d\pi R_p^2}{4t\Omega d^2},
\end{equation}
\citep{worthen24} where $M_d$ is the dust mass, $R_p$ is the planet radius, $t$ is the time it takes the dust in each radius bin to reach the planet from its origin location, $d$ is the distance between the dust origin radius and the planet semi-major axis, and $\Omega$ is the solid angle subtended by the disk. From our spectral fitting, we estimate $M_d$ $\geq$ 3.2$\times$10$^{-3}$ $M_{moon}$ and $t$ $\sim$ 20 years. Detailed modeling of the GRAVITY observations of $\beta$ Pic c and b suggest that their radii are 1.2 and 1.36 $R_{Jup}$, respectively \citep{nowak20a, nowak20b}. If $\Omega$ = 1.92 sr \citep{worthen24}, we estimate average dust accretion rates of $\sim$3 $\times$ 10$^{-16}$ M$_{Jup}$ yr$^{-1}$ and $\sim$10$^{-17}$ M$_{Jup}$ yr$^{-1}$ for $\beta$ Pic c and b, respectively, as the dust cloud swept through the system. These accretion rates are similar to the ones estimated by \cite{worthen24} and approximately one to ten million times lower than those estimated for PDS 70 c and b.

As already described in \cite{worthen24}, the estimated mass accretion rates are very small. Models for accretion of gas and dust in protoplanetary disks onto Jovian mass planets suggest that accretion rates $>$10$^{-9}$ $M_{Jup}$ yr$^{-1}$ are required alter the strength of the continuum emission at 2.5 - 5 $\mu$m (P. Gao, private communication). This suggests that the corresponding debris disk dust accretion rate needed to create a continuum signature, $>$10$^{-11}$ $M_{Jup}$ yr$^{-1}$, is still one million times higher than implied by the recent collisional event. However, there is still some possibility that the elevated dust accretion could be detected. If the particle are sufficient small, then they may remain aloft in a planet's atmosphere and contribute to silicate clouds. In this case, time monitoring of the 10 $\mu$m silicate feature before, during, and after a collisional event could provide direct insight into planetary dust accretion at the late stages of planetary system formation and evolution. 

\section{Conclusions}
We have extracted a \emph{JWST} MIRI MRS spectrum of the $\beta$ Pic debris disk centered on the host star using a large aperture consistent with apertures used to extract spectra from \emph{Spitzer} IRS observations. We compared the newly extracted \emph{JWST} MIRI spectrum, obtained in 2023, with the \emph{Spitzer} IRS spectrum and find:

\begin{enumerate}
    \item The infrared continuum at 5 - 15 $\mu$m observed in 2004-5 has disappeared. This excess was predominantly generated by hot 600 K dust. If analogous to the hot excess observed in 2023, then the dust was probably located within a few au of the host star.
    \item The 18 and 23 $\mu$m crystalline forsterite features observed in 2004-5 have disappeared. Detailed modeling of the spectrum indicates that the mass of the cold $\sim$100 K crystalline forsterite component by decreased by a factor $\geq$60.
    \item The amplitude and the shape of the 10 $\mu$m silicate emission feature has changed and now declines at slightly longer wavelengths. Detailed modeling of the spectrum indicates that the shift in the 10 $\mu$m feature can be reproduced using 551 K crystalline forsterite optical constants rather than the 300 K crystalline optical constants used to originally model the 2004 spectrum.
    \item Small dust grains in the Rayleigh Limit are expected to have a residence time of a couple to a few decades. Thus, the disappearance of the $\sim$600 K continuum and the 18 and 23 $\mu$m features may be explained by radiation pressure blow out between the two epochs. Therefore, a large collision likely took place in the $\beta$ Pic disk within a handful of years prior to the 2004 observation.
    \item If the small grains were originally created by collisions in the inner 10 au, then a wave a dust would have blown past the $\beta$ Pic b companion, consistent with an instantaneous dust accretion rate $\sim$10$^{-17}$ M$_{Jup}$ yr$^{-1}$. This accretion rate is similar to the 2-3$\times$10$^{-17}$ M$_{Jup}$ yr$^{-1}$ rate estimated by \cite{worthen24} and may have been high enough to alter the appearance of the planetary atmosphere.
\end{enumerate}

We thank our anonymous referee for their helpful suggestions for improving the clarity of our manuscript, Y. Argyriou and P. Patapis for helpful discussions about MIRI MRS performance and data analysis, H. Meng for helpful discussions about Subaru COMICS observations of $\beta$ Pic, and A. Meisner for helpful discussions about the \emph{WISE/NEOWISE} UnTimely Catalog.
CHC and KW acknowledge support from the STScI Director’s Research Fund (DRF) and the NASA FINESST program. This work is supported by the National Aeronautics and Space Administration under Grant No. 80NSSC22K1752 issued through the Mission Directorate. 

\software{
This research has made use of the following software projects:
    \href{https://astropy.org/}{Astropy} \citep{astropy13, astropy18, astropy22}, \href{https://matplotlib.org/}{Matplotlib} \citep{matplotlib07}, \href{http://www.numpy.org/}{NumPy} and \href{https://scipy.org/}{SciPy} \citep{numpy07},
    and the NASA's Astrophysics Data System.
}


\bibliographystyle{aasjournal}
\bibliography{project/references}






\end{document}